# Kidemonas: The Silent Guardian

Rudra Prasad Baksi
Dept. of Computer Science and Engineering
University at Buffalo, SUNY
Buffalo, NY, USA
rudrapra@buffalo.edu

Shambhu J. Upadhyaya
Dept. of Computer Science and Engineering
University at Buffalo, SUNY
Buffalo, NY, USA
shambhu@buffalo.edu

*Abstract—* Advanced Persistent Threats or APTs are big challenges to the security of government organizations or industry systems. These threats may result in stealth attacks, but if the attack is confronted before the attacker end goal has been achieved, the attackers could become aggressive by changing the mode of attack or by resorting to some form of contingency plan, which might cause unexpected damage. Therefore, the attack detection and the notification to the system administrator should be done surreptitiously. This paper presents an architecture, called Kidemonas, to silently detect the threat and secretly report it to the user or the system administrator. This way the attacker is deceived into carrying out the attack, without sending any clear signal so that the defender can buy time to develop countermeasures to deal with the attack. We consider several attack scenarios and perform a security analysis to demonstrate the features of Kidemonas.

## I. Introduction

Emerging economies with huge population are experiencing a growth in the Internet penetration among the masses [1]. These economies have seen a steep rise in the smart-phone ownership among the common masses, but advanced economies not only have access to better technologies but also experience higher rates of usage of the technologies [1] [2]. The common man, the industry, the corporate world and the government agencies are heavily dependent on the Internet based technologies. This is a matter of grave concern in terms of cyber-security. Recently many government agencies as well as the industry have experienced a rise in cyber-attacks [3] [4].

The cyber-attacks are known to have been staged for financial gains as well as to steal sensitive information, and perform disruptive functions in the system or manipulate the outcome of sensitive processes. There can be different types of attacks and threats. This paper looks into a particular type of threat known as the Advanced Persistent Threat or APT.

Advanced Persistent Threat (APT) is a type of stealthy attack wherein first the attacker stealthily infiltrates the system, gathers information, performs reconnaissance missions, and once ample information is available and needed access privilege has been achieved, it carries out the attack. APT campaigns are carried out against companies, government agencies and even individuals. APTs are often targeted attacks, may use state-of-the-art technologies, and if not taken care of, they can maintain privileged access to the system for long-terms causing further damage to the system. The attacker can be anyone, any individual, part of an organization (some of which are working outside legal boundaries), or nation state actors. Sometimes these aforementioned attackers work in isolation and at other times they collaborate. With the easy availability of Malware-as-a-Service (MaaS) platforms, these attackers are able to use sophisticated techniques to mount targeted attacks [8] [14].

Detecting an APT in a system is a daunting task. APTs infiltrate stealthily and remain dormant till they deem it suitable enough to carry-out the task. Sometimes it takes days to years before an attack is discovered. Once an APT infiltrates the system, it installs backdoors to communicate with the command and control (C&C) centers via TCP port 443 [15]. If the attacker figures out that the attack has been detected then the attacker might become more aggressive or change the mode of attack or resort to some form of a contingency plan, which might cause unexpected damage. This paper, therefore, presents an architecture for silently detecting the infiltration of an APT malware and to secretly report it to the user or the system administrator as a first layer of defense. Our architecture, called Kidemonas, makes use of a hardware resource called the Trusted Platform Module or TPM. TPM provides a Trusted Execution Environment within which an APT malware detection system can be implemented. Once an APT malware is detected, TPM secretly reports it to the system administrator and the attacker is deceived into carrying out the attack without knowing that the detection has taken place. The contribution of this paper is an architecture that uses deception to provide ample time to the defender to come up with a countermeasure without alarming the attacker who might resort to dangerous contingency plans if discovered.

This paper is organized as follows. Section II of the paper gives a background of the hardware resource, TPM. Section III of the paper discusses the threat model used in this paper. Section IV describes the Kidemonas architecture. Section V analyzes the security of the architecture. Finally, Section VI concludes the paper.

## II. Background

TPM or Trusted Platform Module is a hardware module built along the lines of the specifications laid down by the security group, the Trusted Computing Group [9]. TPM renders quite a few cryptographic capabilities to the system, including the generation of cryptographic keys, random number generation and the safe keeping of the user's cryptographic credentials. Currently the main uses of TPM include platform integrity, safe keeping of the keys used for disk encryption and password protection [10]. TPMs are also used in digital rights management, protection and enforcement of software license, and in the prevention of cheating on online activities by providing remote authentication capabilities.

The general specifications, which are laid down, by the Trusted Computing Group (TCG) for TPM are generally meant for the implementation of the hardware-based TPM. But the industry has also tried implementing TPM as a software only module, more or less having the similar capabilities. The actual architecture of TPM is a bit more complex but Figure 1 shows a simplified schema of TPM [7].

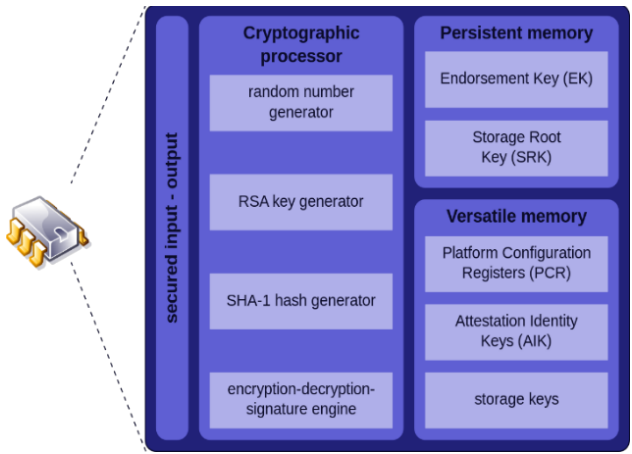

*Figure 1: A simplified Schema of TPM [7]*

The cryptographic processor consists of four main units as shown in Figure 1. The random number generator is used for different purposes depending upon the user process. The RSA key generator produces private and public RSA key pairs for different sessions. If any user process demands hashing, the SHA-1 hash generator produces the hash for the process. Encryption-Decryption signature engine produces RSA digital signatures. TPM, broadly speaking, has two memory units, namely, Persistent Memory and Versatile Memory. Endorsement Key (EK) is a pair of public-private keys which are created and permanently embedded in the memory by the vendor at the time of manufacture. The Storage Root Key (SRK) is a pair of public-private keys invoked when a user takes ownership of the TPM. The SRK is unique to each user and is used for user authentication. If a user gives up the ownership of a TPM then he/she flushes the SRK and a new SRK is generated for the new user who acquires the ownership of the TPM. The SRK and EK are stored inside the Persistent Memory.

Now coming to the Versatile Memory part, as shown in Figure 1, it has the Platform Configuration Registers (PCRs), Attestation Identity Keys (AIK) and the Storage Keys. PCRs are 16-bit registers which store the configuration of the platform. Any deviation from the known configuration indicates a compromise in the platform integrity. The Endorsement Keys are used for platform authentication. The Public Key part of the Endorsement Key is available to all but the Private Part of the Endorsement Key never leaves the TPM and thus can be used for authenticating the platform. The SRK are used by the user to authorize any activity that is taking place inside the TPM.

Intel Software Guard Extension (Intel SGX) and ARM TrustZone, take the aforementioned capabilities of TPM a step further. They provide a Trusted Execution Environment [11] [12] [13]. This environment which Intel names Enclave [11], is inaccessible to any instructions from outside, not even the high priority OS instructions running at the kernel have access to it. Only the processes authorized by the user are allowed to run in the trusted environment. Most of the devices today are available with an in-built TPM unit. But the TPMs in different devices come in different incarnations, depending on the manufacturer producing them. For example, in late 2015, Intel introduced SGX in their sixth generation core microprocessors based on their Skylake microarchitecture [31].

TPM is generally used to check for platform integrity. It also provides cryptographic capabilities. But its usage has been mostly passive. This paper leverages the capabilities of the TPM and the concept of the Trusted Execution Environment to design the hardware-based security system and the deception mechanism.

III. THREAT MODEL

An APT malware doesn't carry-out the attack right away after the infiltration but the attack typically takes place in multiple stages. The "Cyber Kill Chain" model, which has been patented by Lockheed Martin, gives an idea about the life cycle of an APT attack [16]. The APT life cycle conceived by Lockheed Martin consists of 7 stages. LogRhythm on the other hand considers a 5 stage APT life cycle [17] and Lancaster University considers a 3 stage APT life cycle [18].

Considering the aforementioned APT life cycles, and reports from [20] and the model used in [19], we summarize and consider the following APT life-cycle for the Kidemonas Architecture.

- *Initial Reconnaissance and Compromise*: The attacker gathers information regarding the target and stealthily infiltrates by exploiting zero-day vulnerabilities, through social engineering, phishing, malware on frequently visited web-pages, etc.
- *Establish foothold*: In this stage the attacker installs backdoors, Trojans, etc., to communicate with the command and control (C&C) centers.
- *Escalate privileges*: In this stage the attacker gains administrative or root privileges, install rootkits for long-term accesses, and gains ample knowledge for password cracking.
- *Internal reconnaissance*: In this stage the attacker collects internal information and communicates them to the command and control (C&C) centers.
- *Move laterally*: In this stage the attacker searches for and compromises more internal systems.

- *Maintain presence*: In this stage the attacker surreptitiously maintains a control over the compromised system, without raising an alarm in the system.
- *Complete mission*: In this stage the attacker finally carries out the attack which can be stealing of sensitive information, carry-out disruptive processes, subverting the mission when the time is suitable or resorting to a contingency plan if needed.

Stuxnet is one of the first APT malware that received an attention that spanned worldwide [26]. In the following years, few more APT malwares were discovered, such as Duqu and Flame. They belong to the same family as Stuxnet [26] and are more complex and aggressive in nature as compared to Stuxnet. The authors of [28] believe that Duqu might have been created by the creators of Stuxnet. The authors of [27] analyzed Duqu and came to the conclusion that Duqu heavily reuses the Stuxnet code. Stuxnet scans for Siemens Step7 software in the computers with Windows OS which are used for controlling PLCs. If it discovers the software then would introduce unexpected commands on those software, otherwise it would remain dormant. Duqu, even though it heavily re-uses the Stuxnet source code, looks for information and stealthily steals it. It can be used for specific payload delivery as well. It more aggressively looks for Windows vulnerabilities to achieve the desired results. These discoveries bolster our assumption that discovery of a given APT malware may force the attackers to resort to more aggressive attack models. Having looked into different APT attack models, we consider a dynamic threat model. A typical APT may or may not have a contingency plan. But we assume that the threat model we are dealing with, will always have a contingency plan.

In our model, the APT will begin attack with the "Initial compromise" stage and end with the "Complete mission" stage. After the initial infiltration, the APT malware installs a backdoor to establish a communication channel with the command and control (C&C) centers so as to take orders for further processes to be executed. A typical APT executes the contingency plan in the "Complete mission" stage but this paper assumes an APT, which has the capability to execute the contingency plan, if at all it has one, at any stage, whenever it realizes that it has been detected. The information regarding a detection of the malware is forwarded to the C&C center. The C&C center then might resort to the contingency plan. We assume that if the threat is detected in the very first stage of the attack, then the contingency plan of the attacker is to abort the mission. In the event of attacker being successful in completing the first phase of the mission stealthily, then it goes to complete the second phase of the mission, i.e., installing backdoors and establish a foothold in the system. If the threat is being detected in this stage and the C&C center of the attack is informed about the same, then the contingency plan can be either to abort the mission or to change the mode and the target of the attack so as to camouflage the initial attack. In the later stages, if the attack is being detected then the attacker would not resort to mission abortion, instead it changes the type of attack and/or even the target to become more aggressive and execute the mission quicker than before, giving no time to the defender to deal with the attack.

The aforementioned threat model considered in this paper is more dangerous and difficult to deal with as compared to the threat posed by a typical APT malware attack which comes with no contingency plan. Therefore, we present a deception mechanism to camouflage the reporting of the malware detection information, so that the attacker carries on with the initial mode of attack, without resorting to more aggressive forms of attack.

IV. KIDEMONAS: THE ARCHITECTURE

*A. Trusted Computing: Security in Plain Vanilla Form*

Trusted computing has been thought of as a probable solution to the defense against various cyber-threats [23]. In [22] the authors tried to leverage the functionalities of trusted computing to detect any attack. They relied heavily on constant monitoring of the system and checking the integrity of the platform regularly. In the event of any intrusion being detected, the system raises an alarm. This alarm is also visible to the attacker. The attacker then can change the mode of attack to take some aggressive measures or can resort to some other form of contingency plan for which the defender may not be prepared and the system can suffer unexpected damage. In [21] the authors present a model wherein the TPM is given an added functionality to communicate via the link layer communication channels to other nodes in a networked system. But it also raises an alarm whenever any form of intrusion is being detected and that alarm is visible to all. Therefore considering the implications of the aforementioned threat detection alarms, we present Kidemonas, an architecture to silently detect any threat and surreptitiously reporting it to the system administrator.

*B. Kidemonas*

Deception is an important part of the defense mechanism in cyber-security. The authors of [29] present a software security solution wherein the elements of misdirection and deception are introduced using honey-patches. But our architecture, Kidemonas, uses hardware components, as shown in Figure 2, to introduce the deception scheme. By camouflaging the infiltration detection system, the attacker is deceived into believing that it is successful in maintaining its stealth, and it buys ample time for the defender to come up with countermeasures to thwart the attack. To maintain *the cover*, the traffic is analyzed in a concealed environment. Therefore, an encrypted copy of the incoming traffic is sent to our hardware-based security system, and it is in this unit where the malware detection system analyzes the traffic and surreptitiously reports it to the user or the system administrator. The following sub-sections discuss the working of each unit as the traffic flows through it.

*1) The Firewall*

The first line of defense of a system would be the firewall of the system, which acts as a fence and keeps away the malicious actors from the system. But in an event of the breach of the perimeter or in this case breaching the firewall of the system can pose serious threats to the system. APT malware infiltrates the system stealthily, so it can be assumed that the attacker would be able to pass through the firewall.

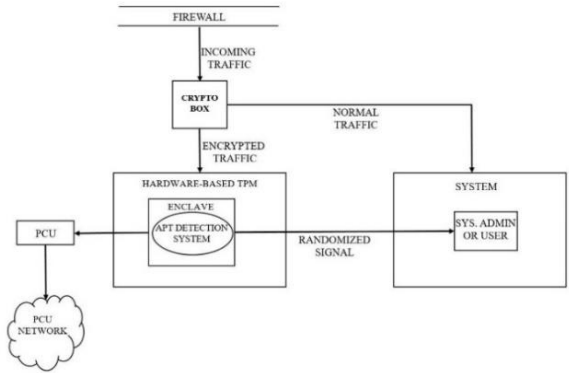

*Figure 2: Kidemonas Architecture*

*2) The Crypto-Box*

After passing through the firewall, the traffic enters the crypto-box, as shown in Figure 2. The crypto-box is used for encrypting the data traffic entering the system. The first task of the crypto-box would be to make a copy of the data traffic entering the system and then to encrypt the copied traffic and sending it to the hardware-based TPM. The crypto-box is configured to encrypt the copied traffic using the RSA-OAEP scheme. The original copy goes to the system as intended.

*3) The Hardware-based TPM*

The encrypted traffic then goes to the hardware-based security system which comprises of a hardware-based TPM unit. This unit, along with various cryptographic capabilities also provides an isolated and exclusive execution environment or the enclave, which even the high priority OS instructions do not have access without user authentication. The idea is to implement the APT Detection System within this execution environment. Any form of infiltration detected by the system, which is running inside the enclave can be communicated to other nodes in the network using the Peer Communication Units (PCUs) as shown in Figure 2. The encrypted traffic can be stored outside the system to further analyze the malware or the attack so that security patches can be produced at the earliest.

Any traffic going inside the TPM doesn't come-out and it behaves like a black-hole. The encrypted traffic entering the TPM, is decrypted using the private key of the TPM. For the hardware-based TPM, the security heavily depends on how securely the private keys are kept hidden. The APT Detection System running inside the TPM analyzes the received traffic and detects any form of infiltration by the APT malware, if it has taken place.

*4) The APT Detection System*

The architecture has been designed in a way that any state-of-the-art APT malware detection system, which is compatible with the system, can be installed within the concealed environment of the hardware-based TPM. Security Information Event Management (SIEM) and User Behavior Analytics (UBA) are popular APT malware detection schemes [14]. The SIEM scheme collects information from different events occurring in different components of the system, and looks for security flaws. This scheme might *blow the cover* of the deception mechanism. But any scheme which analyzes the incoming traffic, like big-data analytics [30], can be installed in Kidemonas for APT malware detection.

*5) The PCU Network*

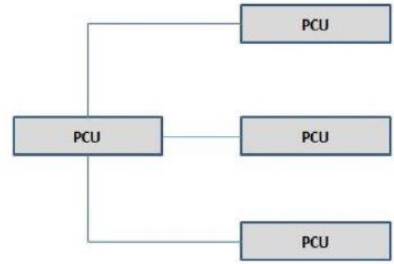

*Figure 3: Peer Communication Unit [21]*

Peer Communication Units (PCU) [21] are the Link-layer or Layer 2 units, which interact with other nodes in the systems via Layer 2. The PCU of one node interacts with the PCU of the other, forming a PCU network as shown in Figure 3. The PCU of a given node interacts only with its hardware-based TPM. Once the APT Detection System, running inside the hardware-based TPM, detects any form of intrusion, it communicates the information to the other nodes in the system through the PCU network. The PCU network actually provides a different communication intra-network link within the system, which is inaccessible to any entity, and is not a part of the system. Even the OS running on each node won't have access to this PCU network.

*6) The Deception Mechanism*

The deception mechanism comprises of an alert, which is covertly raised without letting the attacker know that an alert has been raised. It is achieved by camouflaging the alert as a normal traffic between the TPM and the user or the system administrator. The TPM would send two signals at every interval to the user or the system administrator. The two signals are: Time Stamp and a Random Signal of bit-strings of 0s and 1s. The time stamp will be in the 24-hour format. Since, most of the OS being used today use 64-bit bus, the random bit string will be 64-bits long. At the event of an infiltration being detected, the TPM will generate a time-stamp and use the hour part of the time-stamp as the recognizing bit. For example, if the hour part of the time-stamp is 09, then the bit number 9 of the random string will be set as 1 as well as, (63-9) or the bit number 54 would also be set as 1. In the event if no infiltration being detected and the hour part of the time stamp is 09, then the bit number 9 of the random string will be set as 0 as well as, (63-9) or the bit number 54 would also be set as 0. The user or the system administrator will just check the bit number *n* and the bit number *(63-n)*, as the bits are numbered from 0 to 63, and would realize if there has been any intrusion or not. The infiltration detection is being camouflaged in the randomness of the bit string.

V. ANALYSIS OF THE SECURITY OF KIDEMONAS

*A. The Deception Mechanism*

The most significant aspect of the security provided by Kidemonas is the deception mechanism. Considering the different stages of the APT life cycle (discussed in Section III), the security provided by Kidemonas is analyzed in detail. Kidemonas won't be able to deter the attackers in the first three stages. The attacker will be able to perform initial reconnaissance and infiltration, establish a foot-hold and escalate the privileges. But it won't be able to perform internal reconnaissance, and move and compromise internally as far as the Kidemonas unit is concerned. Each time the infiltrated malware receives any command or information from the C&C centers, Kidemonas would receive the same without the attacker knowing about it. Since, the attacker doesn't know about Kidemonas, and no alarms are being raised on infiltration, it would believe that it has successfully and stealthily compromised the system. Therefore, neither the attacker would become aggressive nor would resort to the contingency plan. This would buy ample time for the defender to come up with a patch or better security features, before the attacker achieves the last stage, i.e., the mission completion stage. This would help the victim to avert any major damage.

*B. The Crypto-Box*

The Crypto-box uses the RSA-OAEP scheme to encrypt the traffic being sent to the hardware-based security unit. It could use the public-key portion of the endorsement key (EK) of the TPM to encrypt the traffic. Now, RSA-OAEP according to [24] and [25] is chosen cipher-text attack (CCA) secure under the RSA Assumption. This would camouflage the copy of the traffic going in to the security unit. Moreover, as discussed earlier, the private key portion of the endorsement key of the TPM never leaves the TPM and the security heavily depends on how the endorsement key is kept secret. Therefore it would not be possible for the attacker to have a better knowledge of the private key than a random guess.

*C. The PCU Network*

Another important aspect of the Kidemonas architecture is the PCU network. This network connects the PCUs of each node in the system. It communicates the information regarding the infiltration of an APT malware to other nodes as well. This communication takes place without the knowledge of the attacker. In a networked system, if one node of the system is able to detect an infiltration, then the entire system is made aware of the same. This helps the entire system to stop the attacker from wreaking havoc and the damages can be minimized to a great extent.

*D. The Trusted Execution Environment*

Last but one of the most significant contributions of the Kidemonas architecture along with the surreptitious reporting scheme is that it also provides a trusted and isolated execution environment wherein any state-of-the-art APT malware detection system can be implemented. With time the APT malwares would become more resourceful and might deceive the existing malware detection systems. The APT malware detection system can be updated manually for newer and more advanced threats. This flexibility of system upgradation makes the architecture robust against becoming obsolete in the face of newer and more advanced form of threats.

*E. The Security Overview in Presence of a Threat*

Now, taking an example of the threat model, the effectiveness of the Kidemonas architecture can be discussed. The threat model perceived in this paper is a dynamic one, which comes with a contingency plan and has the capability to resort to the contingency plan any time it deems necessary and the trigger for such a scenario is the information being conveyed to the attacker that it has been detected by the system. The attacker, if detected in the very first phase, may abort the mission. This doesn't give time to the defender to learn about the attack in details. Kidemonas, surreptitiously informs the system administrator regarding the attack, and the attacker is deceived into carrying out the attack. This gives the defender ample time to study about the attack, during its entire life-cycle and stop the attacker from giving the final blow in the very last stage, as by then the defender would be well armed and ready to thwart the attack. In the second stage, the APT malware installs a backdoor to communicate with the C&C centers of the attack. If discovered in this stage, the attacker can either abort the mission or resort to more aggressive means. Both the scenarios are disadvantageous to the defender. In the former case, it doesn't get ample time to study the attack and in the latter case, it is not ready for a more aggressive form of attack. But, with Kidemonas, camouflaging the infiltration report, the attacker is made to believe that no detection has taken place. This gives defender enough time not only to study about the attack but also about the attacker by studying and analyzing the communication traffic between the malware and the C&C centers. This might also help the defender to be prepared not only for the current form of attack but also for any aggressive contingency plan the attacker might have. Similarly, for the later stages of the attack, the attacker is still deceived into believing that the system is unable to detect its threat and doesn't give the attacker any time to resort to the contingency plan, when the defender thwarts the attack in a well prepared manner in the final stage of the attack.

*F. The Weakness*

But no design is completely secure and each comes with its own share of weaknesses. The biggest concern for the Kidemonas architecture is the insider threat. As of now it is very hard to deal with insider attacks. In a networked system, if the attack happens from one of the nodes, then the entire system is compromised.

## VI. CONCLUSION

We analyze the threat from APT malware and consider an aggressive form of threat model for designing a security system against such an attack. We present Kidemonas, an architecture of a deception mechanism. The future work for this paper would be forming a test-bed and analyzing the architecture in the presence of a real-life threat. It is not always easy to work with real TPMs as different manufacturer come with different specifications and introducing changes at firmware levels as well as OS level would not be very cost effective. So, initially our architecture would be implemented on emulators available commercially for the developers. Once a robust model is prepared, the next step would be to implement the architecture on real hardware. To begin with, it would be a single node system but with time the system will be scaled up to a multiple node networked system. The Kidemonas architecture presented in this paper can also be used for other kinds of threats (subject to testing and analysis). One of the most important application of Kidemonas can be in the field of cloud computing. Kidemonas can be used to take care of the trust issues as well as infiltration from an attacker.

ACKNOWLEDGEMENTS

This research is supported in part by the National Science Foundation under Grant No. DGE – 1241709. Usual disclaimers apply.